\documentclass[aps,twocolumn,showpacs,pre]{revtex4-1}
\usepackage{graphicx,graphics}
\usepackage{amsmath,amssymb,amsfonts}
\usepackage{bbm,bbold}
\usepackage{multirow}
\usepackage{hyperref}
\usepackage{graphicx}
\usepackage{color}

\newcommand{\tw}{\textwidth}

\begin{document}
\title{A simple model of epidemic dynamics with memory effects}

\author{\bf Michael Bestehorn}
\affiliation{Brandenburgische Technische Universit\"at Cottbus-Senftenberg,
  Institut f\"ur Physik, Erich-Weinert-Str. 1, 03046 Cottbus, Germany,
  orcid: 0000-0002-3152-8356 \\
  bestehorn@b-tu.de}

\author{\bf Thomas M. Michelitsch$^*$, Bernard A. Collet}
\affiliation{Sorbonne Universit\'e, Institut Jean le Rond d'Alembert, 
CNRS UMR 7190, 4 place Jussieu, 75252 Paris cedex 05, France,  
$^*$orcid: 0000-0001-7955-6666 \\
thomas.michelitsch@sorbonne-universite.fr, bernard.collet@sorbonne-universite.fr}

\author{\bf Alejandro P. Riascos}
\affiliation{Instituto de F\'isica, Universidad Nacional Aut\'onoma de M\'exico,
Apartado Postal 20-364, 01000 Ciudad de M\'exico, M\'exico,
orcid: 0000-0002-9243-3246 \\ aperezr@fisica.unam.mx}

\author{\bf Andrzej F. Nowakowski}
\affiliation{Department of Mechanical Engineering, University of Sheffield,
Mappin Street, Sheffield S1 3JD, United Kingdom,
orcid: 0000-0002-5018-2661 \\
a.f.nowakowski@sheffield.ac.uk}

\begin{abstract}

\noindent
We introduce a modified SIR model with memory for the dynamics of epidemic spreading in a
constant population of individuals. Each individual is in one of the states susceptible
(${\bf S}$), infected (${\bf I}$) or recovered (${\bf R}$). In the state ${\bf R}$ an
individual is assumed to stay immune within a finite time interval.
In the first part, we introduce a random life time or duration of immunity which is drawn
from a certain probability density function. Once the time of immunity is elapsed an
individual makes an instantaneous transition to the susceptible state. By introducing
a random duration of immunity a memory effect is introduced into the process which
crucially determines the epidemic dynamics. In the second part, we investigate the influence of the memory effect on the space-time dynamics of the
epidemic spreading by implementing this approach into computer simulations and employ a multiple random walker's model. If a susceptible walker meets an infectious one on
the same site, then the susceptible one gets infected with a certain probability.  The computer experiments
allow us to identify relevant parameters for spread or extinction of an epidemic. In
both parts, the finite duration of immunity causes persistent oscillations in the
number of infected individuals with ongoing epidemic activity preventing the
system from relaxation to a steady state solution. Such oscillatory behavior is
supported by real-life observations and cannot be captured by standard SIR models.
\\[3mm]
\end{abstract}

\maketitle

\section{Introduction \label{intro}}

\noindent
The history of mathematical modelling in epidemic spread can be traced back to
Daniel Bernoulli in 1760 \cite{Bernoulli1760}. However, it was much later by the seminal work of Kermack and McKendrick \cite{kermack} that this field became a modern and active
area of research. The basic approach they introduced is the so called 
`SIR model' (${\bf S}=$ susceptible, ${\bf I}=$ infected, ${\bf R}=$ recovered).
It turned out that the dynamics of some infectious diseases such as measles, mumps, and rubella can be well captured in a nonlinear dynamics framework such as SIR type models. For the most simple case of spatially homogeneous infection rates, several versions of SIR models have been introduced \cite{AndersonMay1992,Martcheva2015}.
Among the wide range of SIR type
models we mention here a recent one based on continuous-time random walks
\cite{Angstmann-et-al2021} motivated from fractional dynamics with anomalous transport and diffusion effects \cite{MetzlerKlafter2000,Barkai-et-al2000,SandevChechkinMetzler2021,TMM_Riascos2020,TMM-APR-GFPP2020} which may be important mechanisms in epidemic spreading.
\\[2mm]
It is unsurprising that the interest in this field has literally exploded in the recent years driven by the present
pandemic Covid-19 context. Some related models can be found in the references \cite{BelikGeiselBrockmann2011,Feng-et-al2020}. 
The application of general approaches introduced in epidemic modelling, especially those related to stochastic processes and dynamics indeed have turned out to be fruitful to open a wide new interdisciplinary area of research.
These approaches were further enriched by the emergence of network science with pertinent applications in transportation processes on
complex networks as models for human societies, online networks, transportation networks again have boosted this area as a vast interdisciplinary field. Many of these problems can be described as random walk on complex graphs for which an elaborated framework exists
\cite{WattsStrogatz1998,AlbertBarabsi2002,VanMieghem2011,Barabasi2016,NohRieger2004,Holme2015,Newman2010,Hughes1996,TMM-APR-ISTE2019,RiascosMateos2017} among them various random walk models in complex biased graphs 
\cite{BenziDurastante2020,BianchiDurastanteMazza-et-al2021,ArrigoDurastante2021,RiascosMicheltschPizarro2020,RiascosMateos2021,MichelitschPolitoRiascos2020} to name but a few.
\\[2mm]
Epidemic spreading in complex networks was studied in several works (and many others)
\cite{Satorras-Vespigniany2001,PastorVespigniani2001B,Pastor-Cestellano-Mieghem2015,BesRiascosMichel2020}, among them scale-free networks \cite{PastorVesp2003} and activity-driven adaptive temporal networks \cite{Macastropa-et-al2020} including percolation effects in small-world networks \cite{MooreNewman2000,NewpannWattis1999}. A renormalization group model of the second COVID wave in Europe has been established \cite{CacciapagliaSannino2020}. 
\\[2mm]
Despite of the vast fund of sophisticated models, the variety of newly observed phenomena 
makes it more than ever desirable to develop sufficiently simple models containing a minimal set of parameters to allow identification of the relevant ones governing the epidemic dynamics. This aim was the main source of motivation for the present paper. 
\\[2mm]
Our paper is organized in two principal parts. In the first part we introduce a modified SIR model by taking into account random duration (life times) of immunity following a
prescribed probability density function (PDF). We consider here especially an Erlang PDF which contains two free parameters and turned out to be flexible enough to capture real-life situations consisting by two essential regimes: In one regime the recovered individuals enjoy all a similar time of immunity with a narrow immunity life time PDF. In the other regime the immunity life times are broadly scattered and may differ considerably from one to another recovered individual. For these two regimes the memory effect is studied. Contrary to the standard SIR model the
so modified model exhibits an infinite set of fixed points with non-vanishing numbers of infected individuals. A local analysis shows the
existence of oscillatory instabilities for certain fixed points, a behavior also
known from delay-differential equations like the Hutchinson model \cite{hutch,besdel}.
The full nonlinear solution for these cases reveals the existence of limit cycles with persistent oscillations in the numbers of infected individuals. In these situations the epidemic activity never
ends, thus herd immunity is not any more well defined. The epidemic dynamics then is characterized by recurrent diminution and outbreaks of the epidemic activity. 
The resulting persistent epidemic activity is in contrast to the standard SIR model where
the disease extincts when herd immunity is reached. 
\\[2mm]
In section \ref{multi_walkers} we apply a multiple random walker's model (see \cite{RiascosSanders2020,BesRiascosMichel2020} for details and the references therein) with a constant population of SIR walkers (where each walker is in one of the states ${\bf S}$, ${\bf I}$, ${\bf R}$) navigating independently on an undirected connected graph. We implement this approach into computer simulations and consider walks on small world 2D lattices where the following infection rule applies.
If a susceptible walker meets an infectious one on the same node then the susceptible walker gets infected with a certain probability. Then we employ the same assumption on the occurrence of a random life time of immunity as in the first part
and simulate this behavior by an Erlang PDF.
We perform a series of computer experiments and identify pertinent parameters responsible for
the spreading, oscillation, or extinction of the epidemic activity.

\section{Modified SIR model with memory}
\subsection{The standard SIR model}
\noindent
Let us briefly recall the standard SIR model \cite{kermack}.
This model considers a population of individuals where each individual is in one of the following
three compartments: susceptible (${\bf S}$), infected (${\bf I}$), and recovered
($=$ immune) (${\bf R}$). We use the notation
$s(t), j(t),r(t) \in [0,1]$ for the fractions of susceptible, infectious and recovered individuals, respectively. Recovered individuals are assumed to be immune for a certain random time which will be specified hereafter. Neglecting all birth and death rates we have a constant population $s(t)+j(t)+r(t)=1$. The standard SIR model reads 
\begin{subequations}\label{sirs}
  \begin{align}
    \frac{ds}{dt} & = -\beta\,j\,s  \label{sira} \\
    \frac{dj}{dt} & = \beta\,j\,s -\gamma\,j \label{sirb} \\
    \frac{dr}{dt} & = \gamma\,j  \label{sirc} \ ,
   \end{align}
  \end{subequations}
where $\beta$ denotes the infection rate and $1/\gamma$ is the average time of being infectious or the time of healing.
The basic reproduction number is related to $\beta$ and $\gamma$ as follows
\[ R_0 = \frac{\beta}{\gamma}\]
and $R_e=R_0\, s $ indicates the effective reproduction number where $R_e-1$ measures the rate of new infections at time $t$ generated by one case $j=1$ 
(see Eq.\,(\ref{sirs}b)).

\subsection{The extended model}

\begin{figure*}[!ht]
\centerline{\includegraphics[width=0.9\tw]{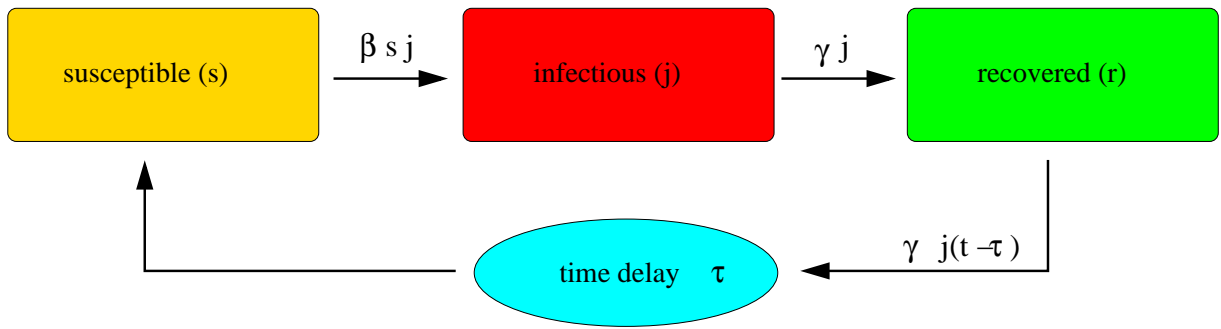}}
\caption{\label{fig1} The standard SIR model is extended by the
feedback loop including a time delay.}
\end{figure*}

\noindent
Now we introduce a generalization of standard SIR where we maintain the assumption of a constant population  $s(t)+j(t)+r(t)=1$.
Contrary to the standard model, where the epidemic dynamics is characterized by the
pathway of the transition ${\bf S} \to {\bf I} \to {\bf R}$
ending in a fixed point $j=0,\ s<1/R_0$, we extend the model to allow an additional
transition ${\bf R} \to {\bf S}$, reflecting the often observed phenomenon of a finite life time of immunity starting after healing (or vaccination), 
see Fig. \ref{fig1}.

The balance of the recovered individuals can then be written as $dr/dt =$
birth rate minus death rate, namely 
\begin{equation}
\label{balance_r}
\frac{d}{dt}r(t) = b(t) - d(t) \ ,
\end{equation}
where $b(t)$ indicates the rate of individuals which (instantaneously) recover at time $t$
i.e. making the transition ${\bf I} \to {\bf R}$. The quantity $d(t)$ stands for the rate of
individuals (instantaneously) loosing their immunity at time $t$ undertaking the transition
${\bf R} \to {\bf S}$.
\\[2mm]
Now we connect this balance equation with a finite life time (duration) of immunity (sojourn
time in state ${\bf R}$) and
introduce the causal probability density function (PDF) $K(\tau)$ from which the finite time of immunity is drawn: An individual that is recovered at instant $0$ (`birth of immunity')
looses its immunity at instant $t'$ (`death of immunity')
with probability $K(t'){\rm d}t'$.
Hence the total death rate ${\bf R} \to {\bf S}$ at time $t$ can be written as
\begin{equation}
\label{trans-rate}
d(t)= \int_{-\infty}^{t}K(t-\tau)b(\tau){\rm d}\tau \ ,
\end{equation}
accounting for the complete history of births $b(\tau)$ taking place up to time $t$. 
The life-time of immunity PDF is normalized,
\begin{equation}
\int_{0}^{\infty}K(t){\rm d}t = 1.
\end{equation}
We will specify the PDF $K(t)$ subsequently.
To keep our model simple, we make the assumption that the birth rate of recovered individuals is given by
$b(t)=\gamma j(t)$, as in standard SIR, i.e. the transition rate ${\bf I} \to {\bf R}$
is assumed to depend only on the value of $j(t)$ at instant $t$, i.e. without additional memory.
\\[2mm]
With these remarks we can now establish a modified set of SIR equations with memory
where we rescale the time $t \to \gamma t$ to arrive at the (dimensionless) form 
\begin{subequations}\label{modified_SIR}
\begin{align}
\frac{d}{dt}s(t) & =   -R_0 s(t) j(t)+\int_0^{\infty}K(\tau)j(t-\tau){\rm d}\tau
\label{mod1}\\
  \frac{d}{dt}j(t) & = R_0 s(t)j(t) - j(t) \label{mod2} \\
  \frac{d}{dt}r(t) & = j(t) - \int_0^{\infty}K(\tau)j(t-\tau){\rm d}\tau \label{mod3} \ .
\end{align}
\end{subequations}
We assume for the analysis to follow that these equations hold for all $t\in \mathbb{R}$
for some prescribed values $s,j,r$ at $t=-\infty$.

\subsection{Stationary solutions and linear stability \label{secfx}}
\begin{figure*}[t!]
	\begin{center}
		\includegraphics[width=0.8\tw]{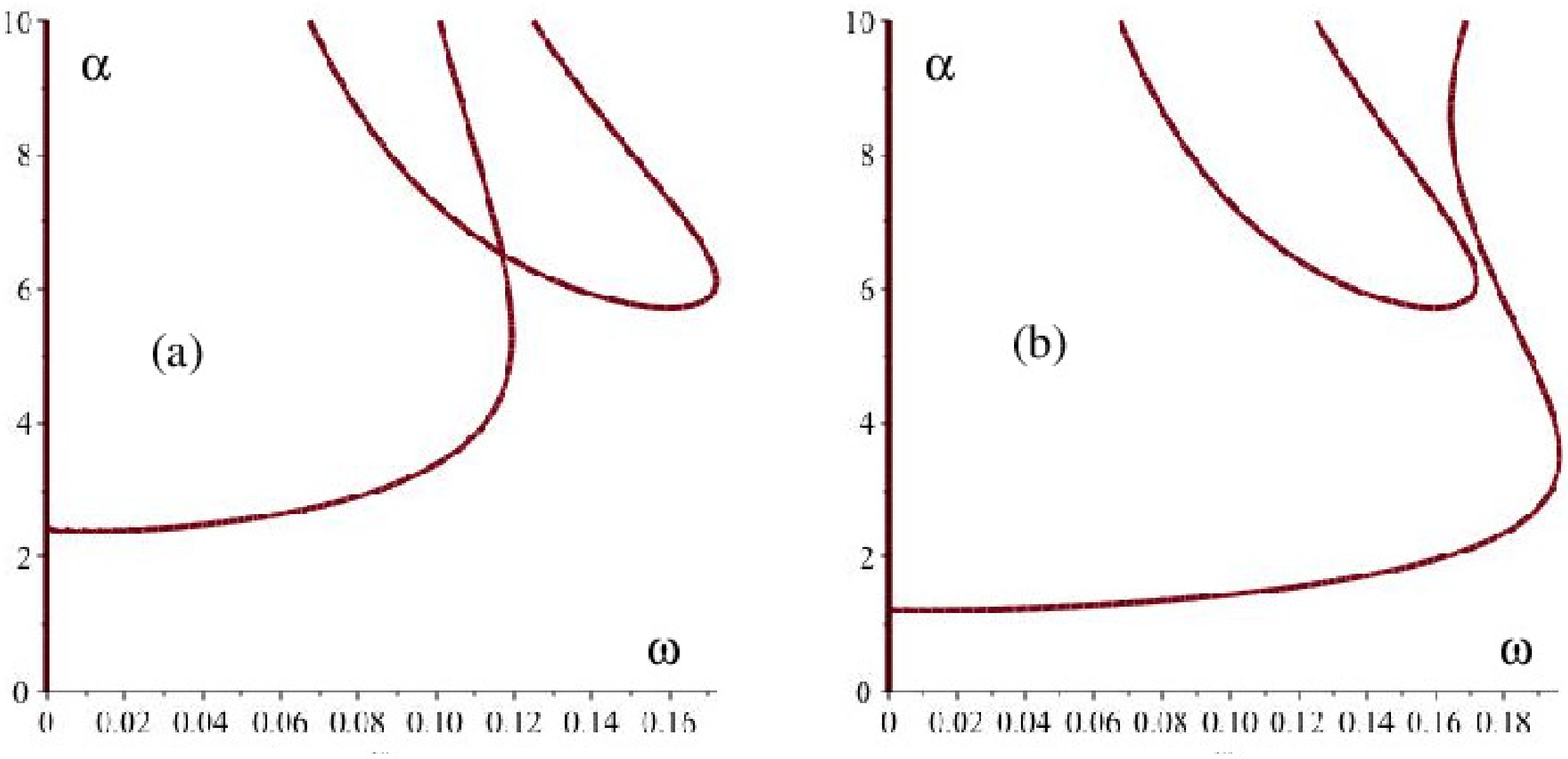}
	\end{center}
	\caption{\label{fig2} (a) The zeros of $f_i$ intersect for $\xi=0.2,\ \varepsilon=0.01$
		at $\omega\approx0.12,\ \alpha\approx6.4$. (b) If $\varepsilon$ exceeds a critical
		value, no solution exists, $\xi=0.2,\ \varepsilon=0.03$}
\end{figure*}
\noindent
The equations (\ref{modified_SIR}) have the following stationary solutions:
\begin{itemize}
 \item $\mbox{(i)}\quad 0\le s_0\le 1,\ j_0=0$,
\item $\mbox{(ii)}\quad s_0=1/R_0,\ 0\le j_0\le 1-s_0$.
\end{itemize}
(i) corresponds to a healthy population with $s_0+r_0=1$ which becomes unstable for $R_e=R_0s_0 \ge 1 $ (outbreak of the epidemic).
\\[2mm]
Linearizing of Eqs. (\ref{modified_SIR}) with respect to the fixed point (ii),
\[ s=s_0 + u\,\mbox{e}^{\lambda t},\quad 
   j=j_0 + v\,\mbox{e}^{\lambda t} \]
yields the solvability condition 
\begin{equation}\label{solv}
  \lambda^2 + \varepsilon\lambda+\varepsilon\left(1 -\hat K(\lambda)\right) = 0 
\end{equation}
where we introduced the abbreviations
\[ \varepsilon = R_0 j_0,\qquad \hat K(\lambda) = \int_0^{\infty} e^{-\lambda t} K(t) {\rm d}t , \hspace{1cm} \Re\{\lambda\} \geq 0 \ . \]
Here, $0\le\varepsilon\le R_0-1$ serves as a bifurcation parameter, $\hat K(\lambda)$
denotes the Laplace transform of the immunity life time PDF and $\Re\{\cdot\}$ stands for the
real part.
\\[2mm]
For an oscillatory (Hopf-) instability with $\lambda=\pm i\omega$, Eq.
(\ref{solv}) turns into
\begin{equation}\label{solv1}
 f_1 =  -\omega^2 + \varepsilon \left(1 -\hat K'(i\omega)\right) = 0, \qquad
 f_2 = \omega -\hat K''(i\omega) = 0 \ ,
\end{equation}
where $\hat K',\ \hat K''$ denote real and imaginary parts of $\hat K$.
At the onset of an oscillatory instability, the two conditions (\ref{solv1})
have to be fulfilled simultaneously.
\section{Immunity life time distribution}
\noindent
In this section we specify the PDF which governs the memory effect by the random life time of immunity of recovered individuals.

\subsection{Erlang distribution \label{secerlang}}

\noindent
An interesting candidate which is able to capture a variety of behaviors
is the so called Erlang distribution (also called gamma-distribution)
which has the form \cite{TMM_Riascos2020} 

\begin{equation}
\label{Erlang}
K_{\alpha,\xi}(t) =  \frac{\xi^{\alpha}t^{\alpha-1}}{\Gamma(\alpha)} e^{-\xi t} , \hspace{0.2cm} \alpha >0, \hspace{0.2cm} \xi >0,\hspace{0.2cm} t\geq 0 \ ,
\end{equation}
where the index $\alpha$ may take any positive (including non-integer) values and
$\Gamma(\alpha)$ denotes the Euler Gamma-function which recovers the standard factorial
$\Gamma(\alpha+1)= \alpha!$ when $\alpha\in \mathbb{N}_0$. For $\alpha=1$ the Erlang distribution
turns into an exponential distribution. The constant $\xi^{-1}$ defines a characteristic time
scale and has physical dimension of time.
For $\alpha \to 0+$ ($\xi$ finite) we have the limit of a Dirac-$\delta$ function $K_{0+,\xi}(t)=\delta(t)$ which 
also is taken for $\alpha$ finite and $\xi \to \infty$.
For $0<\alpha \leq 1$ the Erlang distribution is completely monotonic (CM) with $\frac{d}{dt} K_{\alpha,\xi}(t) <0$ and for $\alpha <1$ weakly singular at $t=0$. For $\alpha >1$ the CM property breaks
down and the Erlang PDF has a maximum at
$t_{\alpha,\xi}=\frac{\alpha-1}{\xi}$. The Erlang PDF has the Fourier (Laplace-) transform 
\begin{equation}
\label{erlangfourier}
{\hat K}_{\alpha,\xi}(i\omega) = \int_{-\infty}^{\infty} e^{-i\omega t} \Theta(t) K_{\alpha}(t) {\rm d}t = 
\frac{\xi^{\alpha}}{(\xi + i\omega)^{\alpha}} \ ,
\end{equation}
where $\Theta(t)$ indicates the Heaviside unit step function which comes into play by
causality. The Erlang PDF has
a finite mean (expected life time of immunity) $\langle t \rangle =\int_0^{\infty} t K_{\alpha,\xi}(t) = \frac{\alpha}{\xi}$, i.e. large $\alpha$ and small $\xi$ increase the duration of immunity of recovered individuals.
\\[2mm]
We point out that the standard SIR model is contained in our extended model as the limiting case
when all recovered individuals have infinite life times of immunity (limit of eternal immunity $\langle t \rangle \to \infty$).
\\[2mm]
For $K(\tau)=K_{\alpha,\xi}(t)$ given by the Erlang PDF (\ref{Erlang}), the system (\ref{solv1}) becomes rather involved. 
A graphical solution is found plotting
the zero lines of $f_i$ for certain fixed values of $\xi$ and $\varepsilon$ in the
$\alpha$-$\omega$ plane and looking for their intersections, Fig. \ref{fig2}.
For later use we point out the following feature of the Erlang PDF allowing a great flexibility to prescribe a globally sharp time of immunity $t_0$ or a broadly scattered distribution.
The possibility to prescribe a sharp expected immunity life time $\tau_0$ is ensured by the limiting property ($\alpha/\xi = \tau_0$)
\begin{equation}
\label{feature_flexible}
\lim_{\xi \to \infty} K_{\xi \tau_0 ,\xi}(t) =\delta(t-\tau_0)
\end{equation}
which is easily confirmed by performing this limit in its Fourier transform ${\hat K}_{\xi \tau_0,\xi}(i\omega)=(1+i\omega/\xi)^{-\xi \tau_0} \to e^{-i\omega \tau_0}$ yielding indeed the Fourier transform of the Dirac's $\delta$-distribution (\ref{feature_flexible}). We consider this case more closely in subsequent section.
\begin{figure*}[t!]
\centerline{\includegraphics[width=0.8\tw]{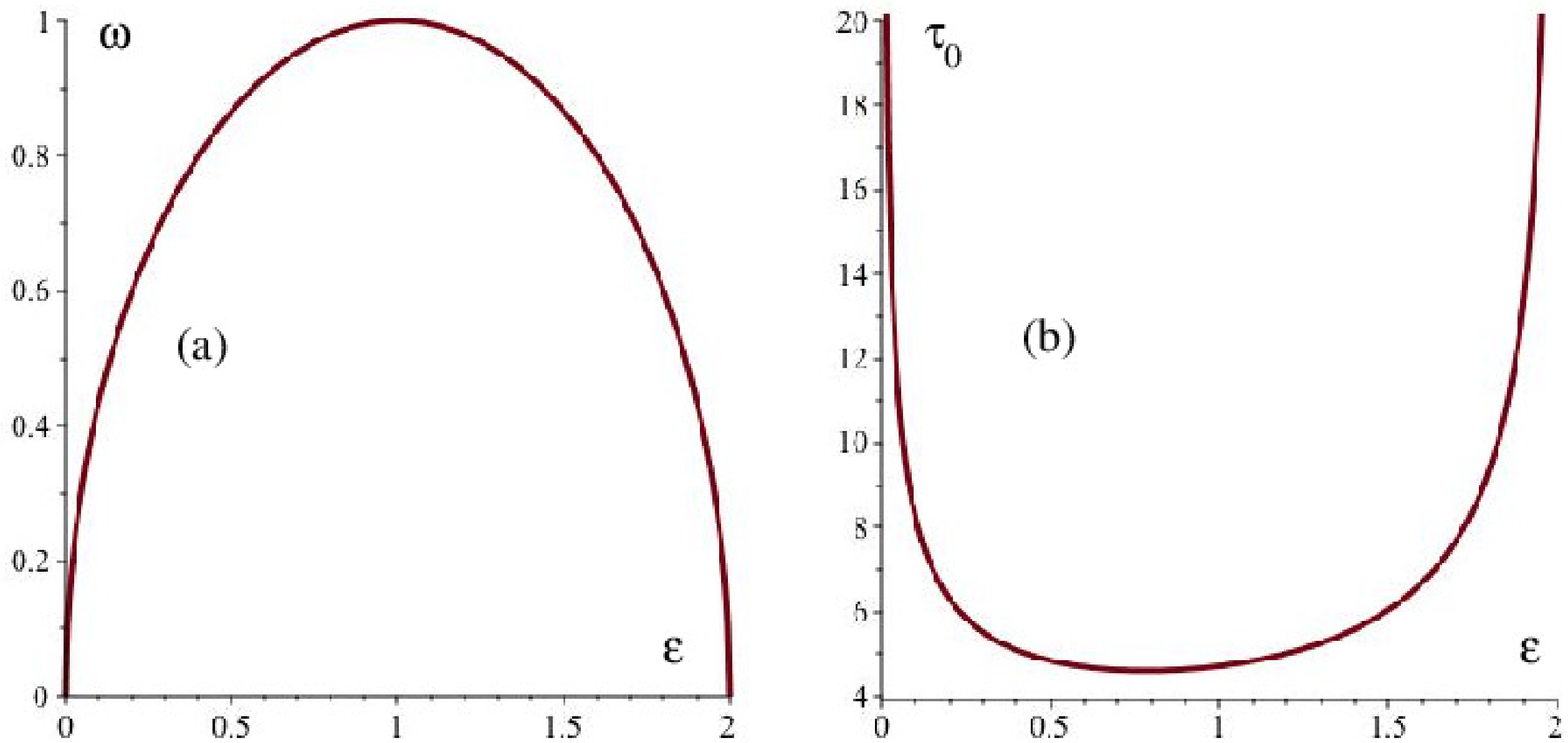}}
\caption{\label{fig3} Frequency (a) and time delay (b) for a delta-distributed kernel.}
\end{figure*}
\subsection{Delta-distribution \label{secdk}}
\noindent
A case that can be evaluated straightforwardly
is that of a $\delta$-distributed kernel $K(t)=\delta(t-\tau_0)$
which is captured by the above limiting case (\ref{feature_flexible}) of the Erlang distribution. The integrals in Eqs. (\ref{modified_SIR}) are then evaluated as
\[ \int_0^{\infty}K(\tau)j(t-\tau){\rm d}\tau = j(t-\tau_0) \ , \]
leading to a set of coupled delay-differential equations (see \cite{besdel1} for a general outline).
Hence Eqs. (\ref{solv1}) take the simple form
\begin{equation}\label{solv2}
 -\omega^2 + \varepsilon \left(1 -\cos(\omega\tau_0)\right) = 0, \qquad
 \omega +\sin(\omega\tau_0) = 0 \ .
\end{equation}
From there one determines (see Fig. \ref{fig3})
\[ \omega=\sqrt{\varepsilon(2-\varepsilon)},\qquad
\tau_0 = \frac{\pi+\arccos(1-\varepsilon)}{\omega} \ . \]
It is clear that also here an upper limit for $\varepsilon$ exists.

\subsection{Numerical solutions}

\noindent
We solved the fully nonlinear system (\ref{modified_SIR}) numerically applying
a standard fourth order Runge-Kutta method \cite{besbuch}. It is sufficient to restrict on Eqs.
(\ref{mod1}), (\ref{mod2}) since $r$ decouples. We used the delta-kernel of
Sec. \ref{secdk}. To evaluate the delay term $j(t-\tau_0)$, the last
$n=\tau_0/\Delta t$ values of $j$ are stored, where $\Delta t$ denotes the
Runge-Kutta time step.
\begin{figure}[b!]
	\centerline{\includegraphics[width=0.45\tw]{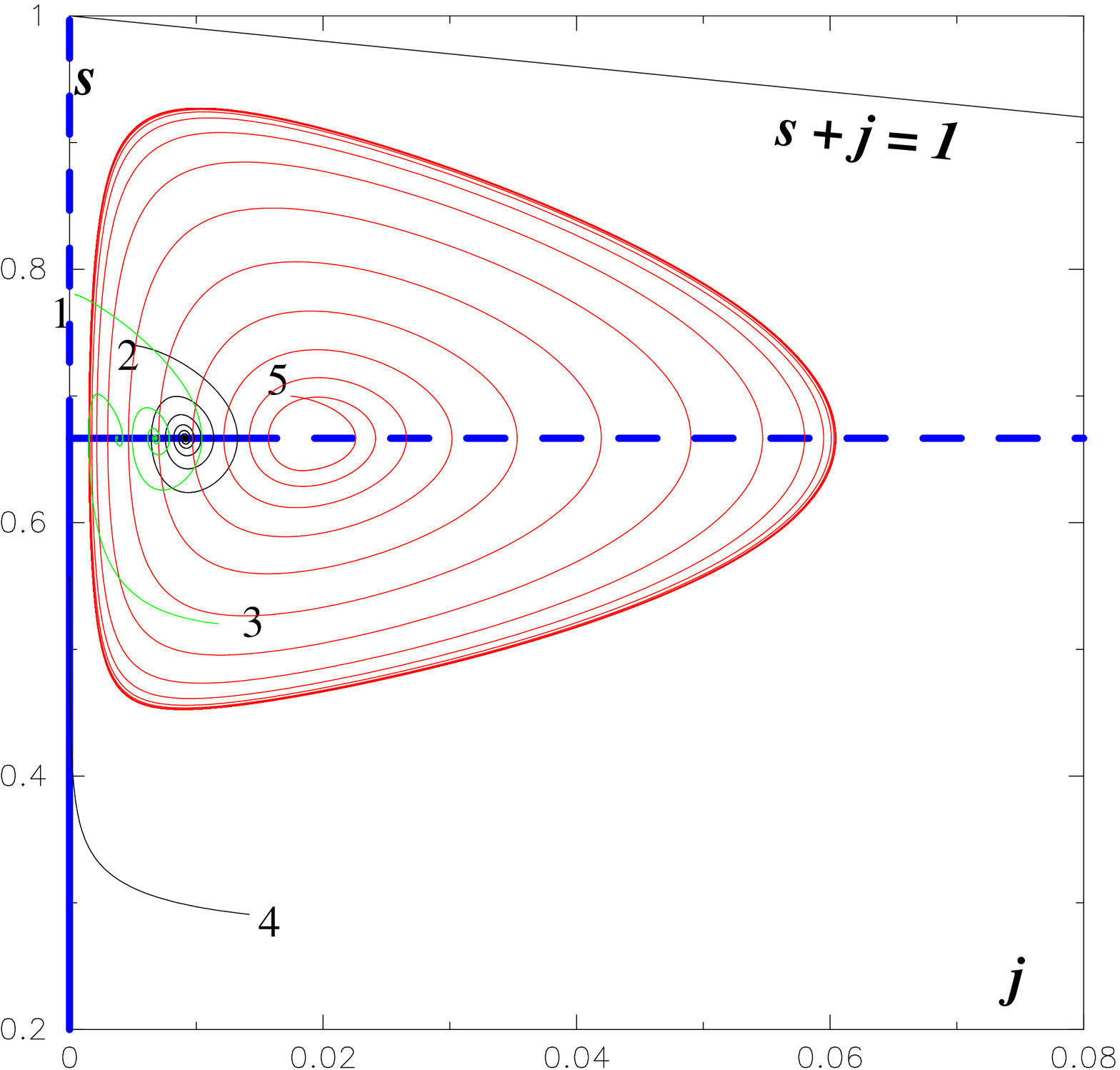}}
	\caption{\label{fig4} Trajectories in the $j$-$s$-phase plane. For different initial
		conditions, different behaviors can be seen. The bold (blue) lines correspond to
		fixed points, solid: stable, dashed: unstable. Starting close to the horizontal
		dashed line, a limit cycle is approached in agreement with the linear computations
		from Sec.\,\ref{secdk}. All trajectories proceed clockwise.
	}
\end{figure}
\\[2mm]
Figure \ref{fig4} shows the $s$-$j$ phase plane. We fixed the basic reproduction
number with $R_0=1.5$.
As initial conditions we use a point $s(0),j(0)$
somewhere in the phase plane and fix the past values of $j$ according to
\[ j(t) = j(0), \qquad -\tau_0\le t\le 0 \ . \]
The value of $\varepsilon$ is then computed from the
initial value $j(0)$ and if $S(0)$ is close to $1/R_0$ the frequency of the
Hopf bifurcation corresponds to that shown in Fig. \ref{fig3}. We chose a
time step of $\Delta t=10^{-4}$, leading to more than $10^5$ iterations per cycle.
The fixed points from Sec.\,\ref{secfx}
are marked in bold (blue), solid for `stable', dashed for `unstable'. The
horizontal dashed line marks the oscillatory instability computed in Sec.\,\ref{secdk},
the vertical one is a monotonic instability. Due to the different local behaviors,
the form of the trajectories depends strongly on the initial condition. For certain
starting points, trajectories may end on a stable fixed point or on a limit
cycle, born at the threshold computed in Fig. \ref{fig3}. However, also the size of the
limit cycle depends on the initial values of $j(0)$ and $s(0)$. For larger values
of $j(0)$ the size of the cycle increases. Note that due to the restriction $j+s+r=1$,
the trajectories must not leave the plane limited by the upper right black line.

\section{2D multiple random walker's approach \label{multi_walkers}}

\noindent
In a previous paper \cite{BesRiascosMichel2020} we considered a population of $Z$ random walkers (particles) to derive estimates for the basic reproduction number
and to explore space-time patterns of the epidemic activity in computer simulations. 
Here we employ the same multiple random walker's model, however, we take into account our above introduced memory effect by assuming a random finite life time of immunity drawn from an Erlang distribution. We also consider limiting cases of a Dirac $\delta$-distribution when the life time of immunity is identical for all 
recovered individuals.

\subsection{The model}

\begin{figure}[t!]
\centerline{\includegraphics[width=0.49\tw]{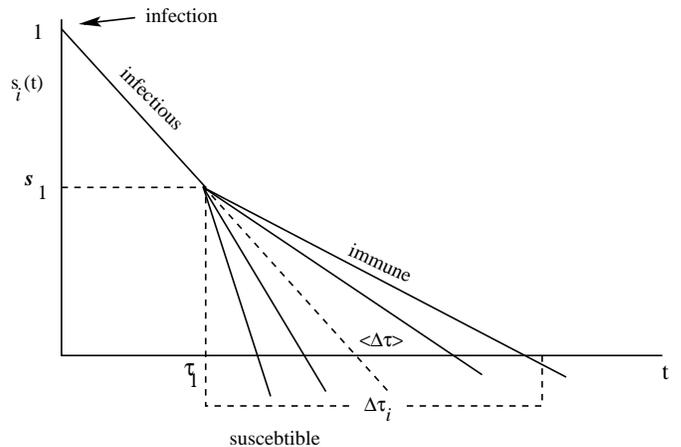}}
\caption{\label{fig5} Linear decrease of $s_i(t)$ after infection at $t=0$ with the identical slope for all infected walkers.
  During immunity ($\tau_1\le t\le\tau_1+\Delta\tau_i$), the slopes are
  individually distributed among the recovered walkers, according to the occurrence of random life time of immunity intervals in our case drawn from an Erlang PDF. Individuals $i$ become again susceptible for $s_i(t)<0$.}
\end{figure}

\noindent
Recall the multiple random walker's model where each walker performs independent jumps at times $t=1,2\ldots$ on a
two-dimensional grid of $N=L^2$ nodes. The positions of the walkers $i=1,\dots,Z$ are indicated by
\[ 1\le x_i^{(n)}\le L, \qquad 1\le y_i^{(n)}\le L \]
where $x_i,y_i,L$ are integer numbers. Here, $n$ denotes the time instants of the jumps.
The walkers may jump according to
\begin{equation}\label{rw1}
  x_i^{(n+1)}  =  x_i^{(n)} + \eta_x^{(n)}, \qquad
  y_i^{(n+1)}  =  y_i^{(n)} + \eta_y^{(n)} \ ,
\end{equation}
with equally distributed random integer numbers $\eta_x,\,\eta_y \in [-h,h]$  where we
consider $h\ll L$ in order to simulate a small-world network. For instance for $h=1$ only jumps
up to the neighbor nodes are possible. Let $s_i^{(n)}$ be an individual state variable characterizing the
`state of health' of walker $i$. If walker $i$ is infected at time $n$, we put $s_i^{(n)}=1$.
To describe gradual recovery effects, we assume a linear decrease in time 
\begin{equation}\label{rw2}
s_i^{(n+1)} = s_i^{(n)} - \mu
\end{equation}
with $1/\mu$ as a global characteristic relaxation time of healing. By choosing the time
step $\Delta t=1$, $s_i^{(n)}$ is synonym for $s_i(n\Delta t)=s_i(n)$.
\\[2mm]
In the present model we assume
for the sake of simplicity that $\mu$ is a global quantity, i.e. identical for all $Z$ walkers.
In other words all infected walkers need the same characteristic time
$\tau_1=(1-s_1)/\mu$ from infection to full recovery (transition ${\bf I} \to {\bf R}$, see Fig.  \ref{fig5}).
\\[2mm]
We define individual $i$ as infectious {\bf I} at time $t=n$ if $1 \geq s_i^{(n)}>s_1$, ({\bf R}) recovered (immune) if $s_1\geq s_i^{(n)} \geq 0$, and ({\bf S}) susceptible if 
$s_i^{(n)}<0$. We depict this behavior of the individual health state variable $s_i^n$ in Fig.  \ref{fig5}.
\\[2mm]
\begin{figure}[b!]

\centerline{\includegraphics[width=0.4\tw]{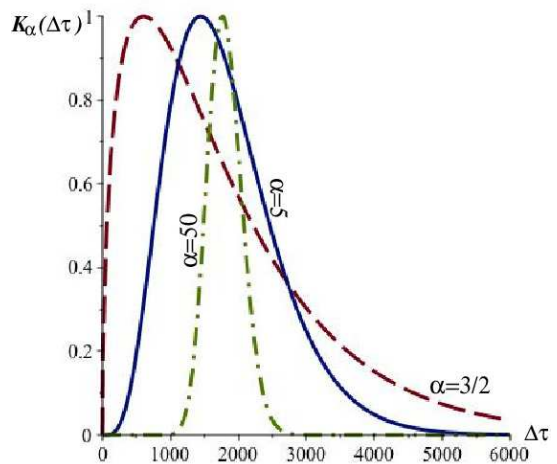}}

\caption{\label{fig6} Scaled Erlang distributions for immunity time $\Delta\tau=1800$ and different 
values of $\alpha$.}
\end{figure}
\begin{figure*}[t!]
	\centerline{\includegraphics[angle=-90,width=0.55\tw]{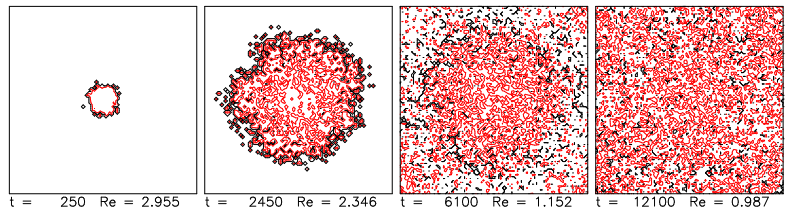}}
	\vspace{2mm}
	\caption{\label{fig7} Time series for initial condition (i), black: susceptible,
		red: infected walkers. For details see text.}
\end{figure*}
\begin{figure*}[t!]
	\centerline{\includegraphics[angle=-90,width=0.55\tw]{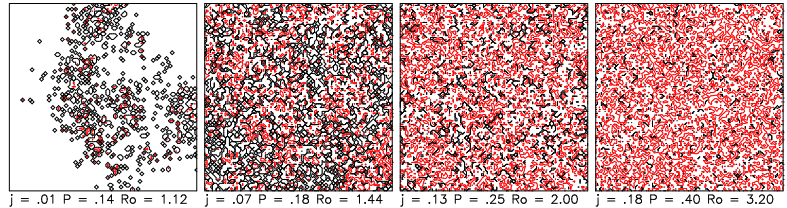}}
	\vspace{2mm}
	\caption{\label{fig8} Long-time distribution for both initial conditions but
		several $P$ and $R_0$.}
\end{figure*}

\noindent
For infection, the following rule applies. If an infected walker $i$ and a susceptible one $j$ meet at the same instant $n$ on the same node, i.e.
\[ x_i^{(n)} = x_j^{(n)},\ y_i^{(n)} = y_j^{(n)} \quad \mbox{and}\quad 
s_i^{(n)} > s_1,\ s_j^{(n)} < 0 \ , \]
then walker $i$ infects walker $j$
with a given probability $P$. In case of infection we reset its state variable $s_j^{(n)} = 1$.
As mentioned we allow here for individual life
times of immunity following a PDF as introduced in Eq. (\ref{trans-rate})
where we focus on Erlang PDF of Eq. (\ref{Erlang}).
Then (\ref{rw2}) takes the more general form, see Fig. \ref{fig5}:
\begin{equation}\label{sfunc}
  s_i^{(n+1)} = \left\{\begin{array}{lcl}
  s_i^{(n)} - \mu, & \mbox{if} & s_1\le s_i^{(n)}\le 1 \\[2ex]
  s_i^{(n)} - \nu_i, & \mbox{if} & s_i^{(n)} < s_1 \ .
  \end{array}\right.
\end{equation}
The individual slopes are given as $\nu_i=s_1/\Delta\tau_i$ (Fig. \ref{fig5}),
where $\Delta\tau_i$ denotes the life time of the immune phase.
\\[2mm]
Let us now specify $\Delta\tau_i$ drawn from an Erlang PDF $K_{\alpha,\xi}(\Delta\tau)$ (see Eq.\,(\ref{Erlang}))
as discussed in Sec.\,\ref{secerlang}, Fig. \ref{fig6}.
Then the values of $s_1$ and $\mu$ can be computed from
\begin{equation}\label{s1mu}
 \mu = \frac{1}{\tau_1+ \langle \Delta \tau \rangle},\qquad s_1 = 1 - \mu\tau_1 
\end{equation}
where $\tau_1$ indicates the time of healing (assumed constant for all individuals) and
\begin{equation}\label{taub}
  \langle \Delta \tau \rangle = \int_0^\infty\tau K_{\alpha,\xi}(\tau)\,{\rm d}\tau = \frac{\alpha}{\xi}
\end{equation}
being the expected (Erlang-) life time of immunity.

\subsection{Numerical results}

\begin{figure}[t!]
\centerline{\includegraphics[width=0.5\tw]{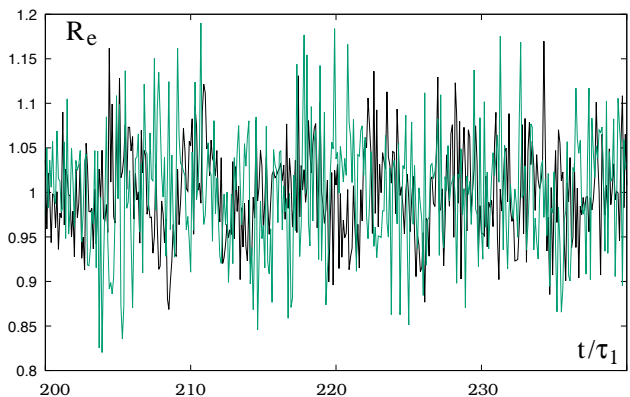}}
\caption{\label{fig9} Effective reproduction number over time for
  $P=0.4$ (black), $P=0.2$ (green).
}
\end{figure}

\noindent
Here we show results on a $N=1500\times1500$ grid with $Z=30000$ walkers and
$\tau_1=600$, $\langle \Delta \tau \rangle = 1800$. The parameters for the Erlang distribution are chosen as
$\alpha=5$ and with Eq.\,(\ref{taub})
$\xi=5/1800$. The basic reproduction number can be estimated as (see \cite{BesRiascosMichel2020} for details),
\[ R_0 = \rho P \tau_1 \ , \]
where $\rho=Z/N\approx0.0133$ is the average density (expected number of walkers on a node). As initial condition 
we assume for the first $Z_I$ walkers being infectious,
\[ s_i^{(0)} = \eta_i, \qquad i=1,\ldots,Z_I \ , \]
where $\eta_i$ are equally distributed random numbers between $s_1$ and 1. The
other walkers are assumed to be healthy and susceptible, 
\[ s_i^{(0)} = 0, \qquad i=Z_I+1,\ldots,Z \ . \]
For the initial positions, we assume (i) all infectious walkers are in the central
position of the grid, and the other (susceptible) ones randomly distributed. (ii) all walkers are
randomly distributed on the grid. For the maximum jump distance of the walkers we take $h=4$.
\\[2mm]
Figure \ref{fig7} shows a time series for (i) with $Z_I=2000$. A dynamics similar to a 
wood fire can be recognized at smaller times. Then the distribution turns into a
more and more random and homogeneous one as long as $R_0>1$. For smaller $R_0$ the
disease extincts. For (ii), the same long time behavior is observed (last
frame in Fig. \ref{fig7}). The mean number of infected walkers depends on the probability of infection $P$ and therefore on $R_0$. This behavior is depicted in Fig. \ref{fig8}.
\\[2mm]
It is interesting to see that the effective basic reproduction number $R_e$
fluctuates around a value of one, quite independently from the probability $P$,
see Fig. \ref{fig9}. 
We compute $R_e$ directly from the simulations by counting the infections per
particle and time step. 
\\[2mm]
\begin{figure}[t!]
\centerline{\includegraphics[width=0.525\tw]{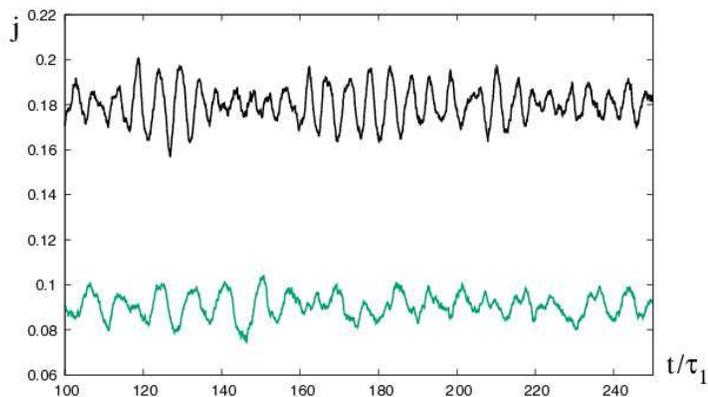}}
\caption{\label{fig10}
Mean relative number of infected walkers over time for
  $P=0.4$ (black), $P=0.2$ (green).
}
\end{figure}
\begin{figure}[h!]
\centerline{\includegraphics[width=0.525\tw]{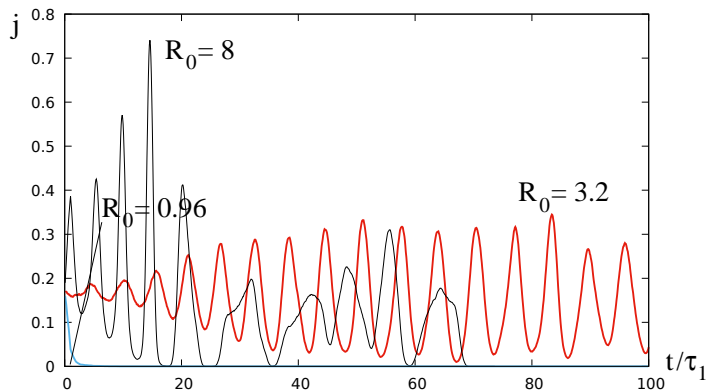}}
\caption{\label{fig11}
  The plot depicts the relative number $j(t)$ of infected walkers vs. $t$ with
  regular oscillations for a delta-kernel (identical immunity life times).
  $j(t)$ oscillates for an intermediate regime
  of $R_0$ (bold, red), where the infection dies out for large $R_0$ (thin) or $R_0<1$ 
  (bold blue).
}
\end{figure}
\begin{figure}[t!]
\centerline{\includegraphics[angle=-90,width=0.475\tw]{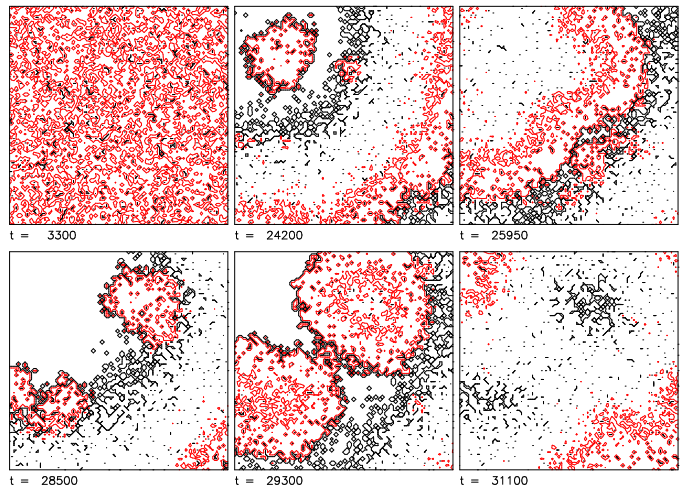}}
\vspace{2mm}
\caption{\label{fig12} Time series for a completely random initial distribution,
  black: susceptible, red: infected walkers. For larger $R_0=6.4$ the infection nearly
  dies out but then spreads again from certain isolated centers. Compare also the rates
  in Fig. \ref{fig11}.}
\end{figure}
\begin{figure*}[t!]
\centerline{\includegraphics[angle=0,width=0.95\tw]{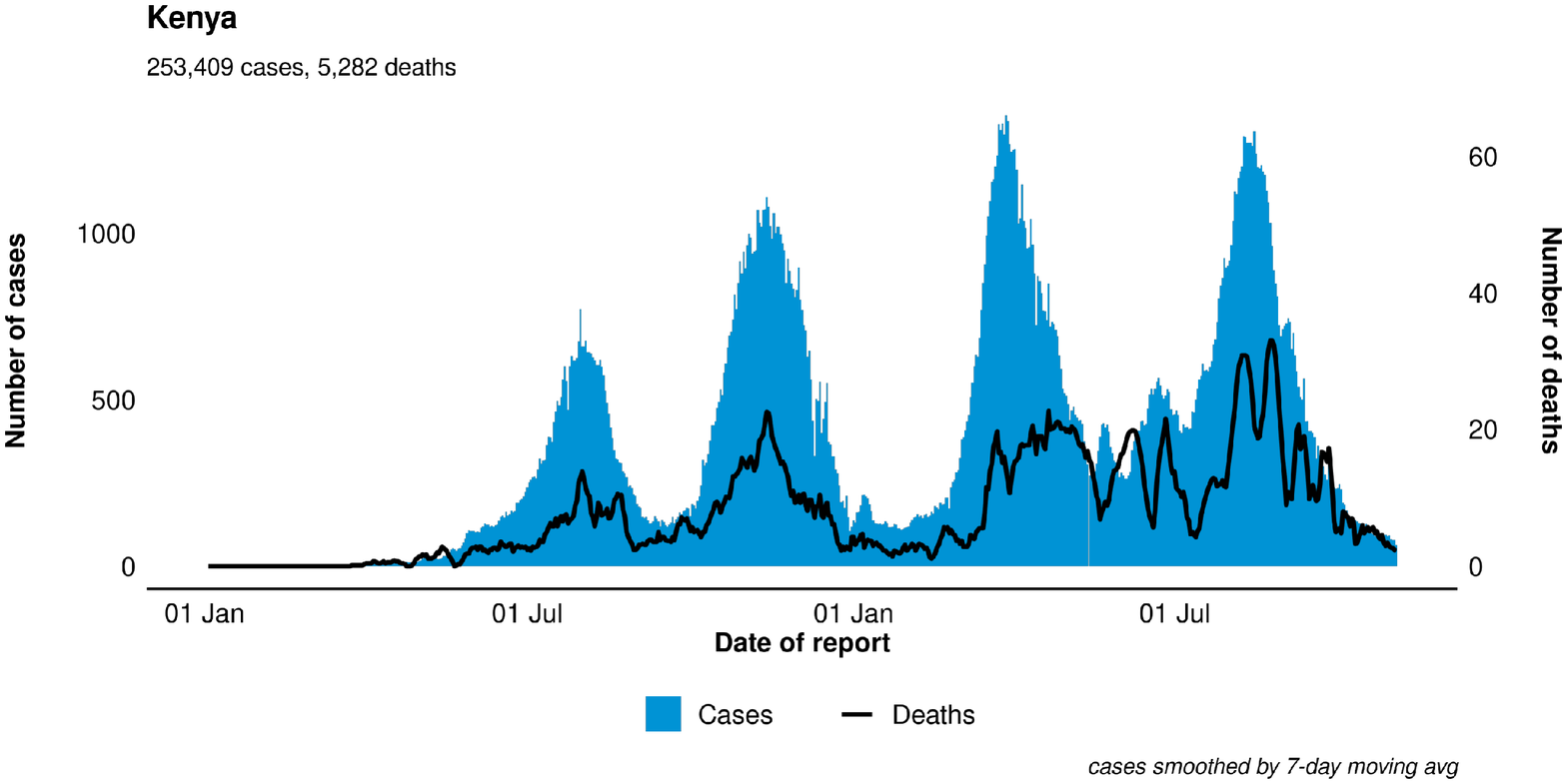}}
\caption{\label{fig13} Time series of the absolute number of cases in Kenya during the years 2021/22 \cite{who}.}
\end{figure*}
\noindent
Another important fact is that the mean numbers of infectious and susceptible
walkers do not asymptotically reach stationary values but rather oscillate
around a mean value with a certain frequency (Fig. \ref{fig10}).
As a consequence the epidemic activity never exhibits extinction at least for $R_0>1$.
This is one of the main differences to the standard SIR model. The standard SIR dynamics
where the epidemics always comes to an end (even for $R_0>1$) is recovered in the limit
$\langle \Delta \tau \rangle \to \infty$ corresponding to infinite life time of immunity.
The oscillatory behavior becomes even more pronounced if the width of the PDF
becomes smaller, i.e. when many individuals have similar immunity life times. In the limit
of a delta-function (all individuals have identical immunity life time), the oscillation become
very regular (Fig. \ref{fig11}) showing synchronization of the walkers for arbitrary
initial conditions. After a certain number of oscillations, the decrease of the relative number of infected individuals reaches almost
extinction but then breaks out again in a certain location and the cycle begins anew. This
behavior can be seen in Fig. \ref{fig12}. These oscillations
exist only in a bounded region of $R_0$. For $R_0<1$, the disease dies out rapidly, for
$R_0$ larger than a critical value that depends also on $\langle\Delta\tau\rangle$
extinction is reached after a certain number of oscillations (Fig. \ref{fig11}).
Qualitatively this is the same scenario found with our extended SIR model, where
limit cycles only exist for $\varepsilon$ below an upper limit.
\\[2mm]
On the other hand oscillatory behavior is supported by the time series of
Covid-19 cases in Kenya for the year 2021, see Fig. \ref{fig13} with recurrent
outbreak of the epidemic activity. Although the observed amplitudes and periods
are different to our model, at least qualitatively an oscillatory epidemic
activity as obtained by our model seems to be supported by these real life data.
Be reminded that such an oscillatory behavior cannot be captured by standard SIR
models.

\section{Conclusions}

\noindent
We proposed an extension of the standard SIR model that considers 
the memory effect introduced by a random
finite immunity time after recovery from infection of the individuals. The
immunity time is supposed to have a certain variation among the individuals and
is described by a PDF, here the Erlang distribution. Contrary to the standard
SIR model, where the disease extincts after one sweep of infection, in our case a regime
of $R_0>1$ may exist with persistent limit cycles leading to a time-periodic behavior of the
number of infectious and susceptible individuals. Depending on the basic reproduction
number $R_0$, the oscillation amplitude of the infected particles can be rather small.
For large $R_0$, the amplitudes may grow in such a way that a kind of ``herd immunity''
is reached at a certain time and the disease extincts.
\\[2mm]
In the second part we considered a multiple random walker's model. It shows
qualitatively the same memory effects: oscillating solutions in an intermediate range 
of $R_0$ whose amplitudes depend on $R_0$, but also on the special form of the PDF
ruling the individual immunity time of the walkers.
The memory effect induces oscillatory characteristics in the epidemic activity where the epidemic activity never ends. This outcome seems to be at least qualitatively supported by real-world situations (Fig. \ref{fig13}). Nevertheless, further quantitative modelling research is needed to confirm this observation.
\\[2mm]
Our model can be extended in different directions. The process of recovery,
i.e. the duration of being ill (infected) can as well be assumed to be random and
modeled by a memory term with another given PDF. On the other hand, spatial
effects can be taken into account considering diffusion terms including space-fractional diffusion with long-range jumps and L\'evy flights \cite{MetzlerKlafter2000,TMM-APR-ISTE2019,RiascosMicheltschPizarro2020}. In
this way, spatially localized structures as encountered in the random walker
simulations may occur. 
\\[2mm]
Further generalizations can be introduced by assuming variable infection probabilities
when susceptible and infected walkers meet. The infection probabilities may vary among the individuals 
and may also depend on time. The interest of such a model is the possibility to capture effects of individually fluctuating virulence, vaccination or resilience to the disease.

\end{document}